\documentclass[a4paper,11pt]{article}
\pdfoutput=1 

\usepackage{jcappub} 

\usepackage[T1]{fontenc} 

\usepackage{cleveref}
\usepackage{natbib}
\usepackage{amssymb}
\usepackage{amsmath}
\usepackage{siunitx}
\usepackage{graphicx,xcolor}
\usepackage[normalem]{ulem}
\usepackage{subcaption}

\usepackage{verbatim}

\newcommand{\bl}{\mathrm{B-L}}
\newcommand{\pbh}{\mathrm{PBH}}
\newcommand{\li}{\mathrm{Li}}

\title{\huge Gravitational Baryogenesis and\\ Dark Matter from Light Black Holes}

\author{Nolan Smyth,}
\author{Lillian Santos-Olmsted,}
\author{and Stefano Profumo}
\affiliation{
    Department of Physics, University of California, Santa Cruz\\
    1156 High Street, Santa Cruz, CA 95064, USA
}
\affiliation{
    Santa Cruz Institute for Particle Physics\\
    1156 High Street, Santa Cruz, CA 95064, USA
}

\emailAdd{nwsmyth@ucsc.edu}
\emailAdd{lsantoso@ucsc.edu}
\emailAdd{profumo@ucsc.edu}

\abstract{We study a scenario in which the baryon asymmetry is created through Hawking radiation from primordial black holes via a dynamically-generated chemical potential. This mechanism can also be used to generate the observed dark matter abundance, regardless of whether or not the black holes fully evaporate. In the case that evaporation ceases, the observed dark matter abundance is generically comprised of both relic black holes and an asymmetric dark matter component. We show that this two-component dark matter scenario can simultaneously account for the observed baryon asymmetry and the cosmological dark matter, a possibility which evades constraints on either individual candidate.}

\begin{document}
\maketitle
\flushbottom

\section{Introduction}
\label{sec:introduction}

The origin of the baryon asymmetry of the universe and the nature of the cosmological dark matter are two of the greatest mysteries facing physicists today. These puzzles lay at the intersection of cosmology and particle physics, leading to significant efforts on both fronts (see e.g. \cite{balazs2014, garcia-bellido2019, cohenSpontaneousBaryogenesis1988, alexander2007}). The baryon-generating mechanism which led to the excess of matter over antimatter implies three conditions be met\footnote{Assuming CPT invariance.}, commonly known as Sakharov Conditions \cite{sakharov1967}:

\begin{enumerate}
    \item Baryon number (B) is violated;
    \item C and CP symmetries are broken;
    \item The process occurs out-of-equilibrium.
\end{enumerate}
The first condition is straightforward; if this condition is not met, there can not be any matter-antimatter asymmetry since baryon number would be conserved automatically. For the second condition, C violation is required, otherwise the process which produces more baryons than antibaryons would be balanced by a corresponding process that produces more antibaryons than baryons. Similarly, CP violation is required so that an equal number of left-handed baryons and right-handed antibaryons are not produced. The third condition follows from (the assumed) CPT invariance; in thermal equilibrium, any process that produces more matter than antimatter will be balanced out by a process that produces more antimatter than matter. 

The idea that the baryon asymmetry is linked to black holes is a natural one; the no-hair theorem requires that the gravitational interaction violate baryon number, implying that the very existence of black holes causes the universe to perform a random walk in B space \cite{semiz1995, semiz2016}. The production of a baryon asymmetry from the evaporation of black holes was first proposed by Hawking \cite{1974Natur.248...30H} and has since been explored further in numerous works. Recent interest in primordial black holes (PBH) as dark matter candidates has led to many studies that consider the prospect of baryogenesis through PBH evaporation. For example, Hawking evaporation into right handed neutrinos or generic GUT scale particles, which subsequently decay via C and CP violating processes, is one such possibility (see e.g. \cite{barrow1991, fujita2014}). Alternatively, the Hawking radiation itself could be asymmetric due to a dynamically generated chemical potential near the surface of the black hole, a processes typically referred to as \textit{Gravitational Baryogenesis}. This was first explored in the context of a coupling between the Ricci scalar and a baryon current, leading to an out-of-equilibrium process which violates C and CP \cite{davoudiaslGravitationalBaryogenesis2004}. This scenario has since been further studied and extended to generic PBH mass spectra \cite{hookBaryogenesisHawkingRadiation2014, boudonBaryogenesisAsymmetricHawking2020}. More recently, the possibility has been raised that a dynamic chemical potential is produced not from the expansion of the universe, but from the time evolution of the black hole itself  \cite{hamadaBaryonAsymmetryPrimordial2017}. This final implementation is the primary mechanism we focus on in the context of a common origin for baryons and DM composed of relic black holes, asymmetric particle DM, or a combination thereof (see e.g. \cite{carr2020, garcia-bellido2019, mcdermott2012, morrison2019}).

Specifically, we consider in this work a chemical potential that is dynamically generated from the time evolution of an evaporating black hole through the term in the action $\cal S$

\begin{equation}
\label{eq:evap}
   \mathcal{S} \supset \frac{1}{M_*^4} \partial_\alpha (\mathcal{R}_{\mu \nu \rho \sigma} \mathcal{R}^{\mu \nu \rho \sigma})J_{\bl}^\alpha,
\end{equation}
where $J_{\bl}^\mu$ is the baryon-lepton current, $\mathcal{R}_{\mu \nu \rho \sigma}$ is the Riemann tensor,  $\mathcal{R}_{\mu \nu \rho \sigma} \mathcal{R}^{\mu \nu \rho \sigma}$ is the Kretschmann scalar, and $M_*$ is a generic energy scale at which the effective operator becomes relevant. This theory violates baryon number, is CP violating, and produces a CPT violating chemical potential, collectively satisfying the Sakharov conditions, as pointed out in \cite{hamadaBaryonAsymmetryPrimordial2017}. Such an interaction may also contribute to the baryon asymmetry through gravitational fluctuations during inflation \cite{alexander2006}. In this work, however, we focus solely on the contribution due to black hole evaporation. 

We note that a lower-dimensional operator can be constructed using the time evolution of the Ricci scalar, $R$,

\begin{equation}
\label{eq:expand}
    \mathcal{S} \supset \sqrt{-g}\lambda \frac{\partial_\mu R}{M_p^2}J^\mu_{\bl},
\end{equation}
where $\lambda$ is a dimensionless coupling constant. In this case, because the Ricci scalar is vanishing for a Schwarzschild black hole, the time evolution is provided through the trace anomaly of the stress-energy tensor for a NFW cosmology (see e.g.
\cite{davoudiaslGravitationalBaryogenesis2004, hookBaryogenesisHawkingRadiation2014}). However, this term requires an enormous coupling ($\lambda \sim 10^{100}$) in order to generate the correct baryon asymmetry \cite{boudonBaryogenesisAsymmetricHawking2020} \footnote{In principle, this large coupling is not a death blow to the theory since one can appeal to our ignorance of a complete theory of quantum gravity. But this requires that no perturbative physics is sensitive to $\lambda$ (see \cite{hookBaryogenesisHawkingRadiation2014}).}. Therefore, we focus exclusively on the coupling in Eq. (\ref{eq:evap}).

We note that the operator in Eq. (\ref{eq:evap}) plays the role of an effective chemical potential, which can be be understood as follows:
When there is a time-dependent curvature, there exists a constant term multiplying $J^{0}_{\bl}$, the baryon-lepton number density charge. Thus, the Lagrangian has a term like $\mu Q_{\bl}$ which behaves exactly like a chemical potential. 

We also note that (assuming $\bl$ is a good symmetry) both (\ref{eq:evap}) and (\ref{eq:expand}) vanish when integrated by parts, up to boundary terms. Thus to all orders in perturbation theory, the above couplings have no effect. However we can either assume the existence of an additional B-violating operator, as in Hamada and Iso \cite{hamadaBaryonAsymmetryPrimordial2017}; Or, as pointed out by Hook \cite{hookBaryogenesisHawkingRadiation2014}, we can use the fact that outside the black hole, an observer can not see beyond the horizon. Thus, one picks up a non-perturbative boundary term due to the black hole and the mechanism works even when $\bl$ symmetry is respected. 

So far, only neutral black holes have been considered in the context of gravitational baryogenesis. However, there is cause for also considering charged black holes. As is well known \citep{hookBaryogenesisHawkingRadiation2014, gibbonsVacuumPolarizationSpontaneous1975}, a charged black hole preferentially emits particles of the same charge, thus losing its charge exponentially quickly. But neutral black holes can also spontaneously emit a charged particle, thus acquiring an electric charge. Therefore, if a black hole is emitting charged particles at a sufficient rate, its net charge will fluctuate around 0 with a root-mean-square value that is non-zero (Q/e $\sim$ 3-6, depending on the mass of the black hole). If evaporation ceases below the Planck scale, it is possible for a black hole to be stuck with a non-zero electric charge and become a charged Planck-scale relic \citep{lehmannDirectDetectionPrimordial2019}. Therefore, we will consider the more general case of a non-rotating black hole with charge $Q$ (although much of the analysis will apply equally well for the case of neutral Schwarzschild black holes).

The form of the initial PBH mass spectrum $\psi(M) \equiv \frac{1}{\overline{\rho_{\mathrm{PBH}}}}\frac{d\rho(M)}{dM}$ corresponds to the epoch during which the PBHs were produced in the early Universe. The mass of a PBH is associated with density fluctuations corresponding to a delta-function power spectrum. It is therefore possible that the mass spectrum is very sharp. In this work, we will consider a delta-function power spectrum for simplicity: $\psi \sim \delta(M-M_{\mathrm{PBH}})$, where $M_{\mathrm{PBH}}$ is the mass of the primordial black holes at formation. 

The reminder of our paper is organized as follows: in \cref{sec:predictions}, we introduce a scenario in which a chemical potential is dynamically generated through a coupling between time-dependent curvature and the Baryon-Lepton current. We show that this can produce the net baryon number observed today and that the remnant black holes can account for the current dark matter abundance. In \cref{sec:Asymmetric}, we explore the possibility of an asymmetric dark matter sector which is produced in a manner analogous to that of baryons. Here, we demonstrate that the asymmetric dark matter can account for 100\% of the dark matter while simultaneously satisfying baryogenesis. In \cref{sec:TwoComponent}, we consider the possibility that the relic abundance of dark matter is comprised of both relic black holes and an asymmetric dark matter particle. We show that such a scenario is possible given current constraints on relic black holes and asymmetric dark matter, again while correctly producing the baryon asymmetry. We conclude with the implications of our results in \cref{sec:discussion}.

\section{Baryogenesis Through Black Hole Evaporation}
\label{sec:predictions}

\subsection{PBH Evaporation}

The energy of a black hole lost per unit area per unit time due to Hawking radiation is given by

\begin{equation}
    \frac{d E}{d x^2 dt} -\sum_i \frac{g_i}{4} \int \frac{d^3 k}{(2\pi)^3} \frac{k}{e^{(k+\mu_i)/T_H} \pm 1} = -\sum_{\mathrm{fermions}} \frac{g_i T_H^4}{4} f(\mu_i) - \sum_{\mathrm{bosons}} \frac{g_i T_H^4}{4} b(\mu_i),
\end{equation}
where $T_H = \frac{\sqrt{G^2 M^2 - G Q^2}}{2\pi (G M + \sqrt{G^2 M^2 - G Q^2})^2}$ is the Hawking temperature of a non-rotating charged black hole, $f(\mu_i) = -\frac{3}{\pi^2} \li_4(-e^{(-\mu_i/T_H)})$, $b(\mu_i) = \frac{3}{\pi^2} \li_4(-e^{(-\mu_i/T_H)})$, and the poly-logarithmic function is defined as $\li_a(z) \equiv \sum_{j=1}^\infty z^j/j^a$ \cite{hookBaryogenesisHawkingRadiation2014}. Note that in the limit $M \gg Q$, which we will assume unless otherwise noted, the Hawking temperature reduces to that of a Schwarzschild black hole, $T = M_p^2/M$, where $M_p$ is the reduced Planck mass. Multiplying by the area of the event horizon gives the rate of energy loss, or equivalently, mass loss. Because the black hole radiation is thermal, this rate can be approximated via the Stephan-Boltzmann law, yielding 

\begin{equation}
\label{eq:dEdt}
    \frac{d E}{dt} = -(8\pi)^2 \alpha \frac{M_p^4}{M^2},
\end{equation}
where $\alpha$ is a numerical coefficient with a value of $3.5 \times 10^{-3}$ from the Standard Model (SM) degrees of freedom \cite{hamadaBaryonAsymmetryPrimordial2017} (Note we are neglecting gray-body factors, which we do not expect to produce a major quantitative effect).
Integrating Eq. (\ref{eq:dEdt}) gives the approximate mass of an evaporating black hole as a function of time 

\begin{equation}
    M(t) = \Big((M_{\pbh}^3 - 3(8\pi)^2 \alpha M_p^4 (t-t_i))^{1/3} - M_f\Big) \Theta[t_f(M_\pbh, t_i) - t] + M_f
\end{equation}
where $\Theta$ is the Heaviside step function, $t_f$ is the time it takes for the black hole to fully evaporate, $t_i$ is the time of PBH formation, and $M_f$ is the final mass of the evaporating black hole. If the black hole completely evaporates, this would correspond to $M_f = 0$. The time at which evaporation ceases (either due to new physics or because of complete evaporation) is thus 

\begin{equation}
    t_f(M_\pbh, t_i) = \frac{M_\pbh^3 - M_f^3}{3(8\pi)^2 \alpha M_p^4} + t_i,
\end{equation}
where $M_f^3$ is always negligible compared to $M_\pbh$ for our cases of interest.

\subsection{Chemical Potential}

A radiating black hole is not a stationary object; a correct description must take into account the energy of the emitted radiation. We therefore consider the outgoing Vaidya-Bonner metric for a charged black hole \cite{vaidya1951, bonnor1970}

\begin{equation}
    ds^2 = \Big( 1 - \frac{2 G M(u)}{r} + \frac{G Q^2}{r^2} \Big) du^2 + 2 dudr - r^2 d\theta^2 - r^2 \sin^2\theta d\phi^2,
\end{equation}
where $u = t-r_*, r_* = r + 2GM \log|(r-2GM)/2GM|$.

The Kretschmann scalar for a non-rotating Vaidya-Bonner black hole is \cite{henryKretschmannScalarKerrNewman2000}

\begin{equation}
    R_{\mu \nu \rho \sigma} R^{\mu \nu \rho \sigma} = G^2 \frac{48 M^2 r^2 - 96 M Q^2 r + 56 Q^4}{r^8}.
\end{equation}
As expected, in the limit that $Q$ goes to $0$, we recover the Kretschmann scalar for a Schwarzschild black hole: $R_{\mu \nu \rho \sigma} R^{\mu \nu \rho \sigma} = 48 G^2 \frac{M^2}{r^6}$ (see e.g. \cite{misner1973}).

The black hole mass evolution, as reviewed above, goes as $\frac{dM}{du} \approx -(8\pi)^2 \frac{M_p^4}{M^2}\alpha$. Therefore, assuming the charge is slowly varying compared to the mass, the corresponding chemical potential is 

\begin{equation}
    \frac{\mu}{T_H} = \frac{3}{2} (8\pi)^6 \alpha \Big(\frac{M_p}{M_*}\Big)^4 \Big(4\pi Q^2 \Big(\frac{M_p}{M}\Big)^8 - \Big(\frac{M_p}{M}\Big)^6\Big).
\end{equation}
where we take $r = r_H$. Note that even if the Hawking radiation originates at $r \approx 2 r_H$, the resulting error is of order 1 \cite{hamadaBaryonAsymmetryPrimordial2017}. For the purposes of this work, we will therefore evaluate the chemical potential at $r_H$. For $M \sim M_p, Q \sim 1$, the charge term is dominant and therefore relevant for our scenario of a significant population of PBH with $M \sim M_p$.

Note that there are two divergences of the radial component of the metric for a charged, non-rotating black hole:

\begin{equation}
    r_{\pm} = G M \pm \sqrt{G^2 M^2 - G Q^2}.
\end{equation}
When $Q \geq \sqrt{G} M$, the black hole is extremal and has no surface gravity. This can be viewed as an upper limit on the charge of a black hole of a given mass, or vice-versa \cite{lehmannDirectDetectionPrimordial2019}. 


\subsection{B-L Charge}
\label{sub:BLCharge}

In this section, we follow Hamada and Iso \cite{hamadaBaryonAsymmetryPrimordial2017} in evaluating the particle asymmetry produced by the evaporation of a black hole. Note that the following calculations assume sphaleron processes are efficient and unsuppressed above the electroweak phase transition, transforming a lepton asymmetry into a baryon asymmetry because of the assumed conservation of $\bl$. 

The average energy emitted per massless particle is

\begin{equation}
    \langle E\rangle = \frac{n_{\mathrm{other}}}{n_{\mathrm{tot}}} \langle E_{\mathrm{other}}\rangle + \frac{n_{L}}{n_{\mathrm{tot}}} \langle E_{L}\rangle + \frac{n_{\overline{L}}}{n_{\mathrm{tot}}} \langle E_{\overline{L}}\rangle,
\end{equation}
where $n_L$, $n_{\overline{L}}$, and $n_{\mathrm{other}}$ refer to the number densities of leptons, anti-leptons, and other SM particles respectively. This labeling scheme is also applied to the average particle energies and will be used throughout. 

The number density of each particle species is given by standard thermodynamic integrals over momenta (see e.g. \cite{kolb1990}):

\begin{align} 
n_{\mathrm{other}} = \frac{g_{\mathrm{other}}}{(2\pi)^3} \int d^3 k (e^{k/T_H}+1)^{-1} = g_{\mathrm{other}} \frac{3 \zeta(3)}{4\pi^2}T_H^3, \\ 
n_{L} = \frac{g_{L}}{(2\pi)^3} \int d^3 k (e^{(k+\mu)/T_H}+1)^{-1} = -\frac{g_L}{\pi^2} T_H^3 \li_3(-e^{-\mu/T_H}),\\
n_{\overline{L}} = \frac{g_{\overline{L}}}{(2\pi)^3} \int d^3 k (e^{(k+\mu)/T_H}+1)^{-1} = -\frac{g_{\overline{L}}}{\pi^2}T_H^3 \li_3(-e^{\mu/T_H}),
\end{align}
where $g$ is the degrees of freedom for each particle species. 

The averaged energy for each particle species is calculated similarly and results in

\begin{align}
    \langle E_{\mathrm{other}} \rangle = \frac{7 \pi^4}{180 \zeta(3)}T_H \\
    \langle E_{L} \rangle = 3T_H \frac{\li_4(-e^{-\mu/T_H})}{\li_3(-e^{-\mu/T_H})}, \\
    \langle E_{\overline{L}} \rangle = 3T_H \frac{\li_4(-e^{\mu/T_H})}{\li_3(e^{-\mu/T_H})}.
\end{align}
The total number of baryons produced per black hole is thus given by

\begin{equation}
    \delta N_{\bl} = \int_{M_f}^{M_\pbh} \frac{dM}{\langle E \rangle} \frac{n_L - n_{\overline{L}}}{n_{\mathrm{tot}}},
\end{equation}
where $n_{\mathrm{tot}} = n_L + n_{\overline{L}} + n_{\mathrm{other}}$.

Note that for sufficiently small $M$, the Hawking temperature $T_H = M_p^2 / M$ eventually becomes larger than $M_*$ (see Eq. \ref{eq:evap}). At this point, the validity of the effective field theory description becomes tenuous and a UV complete theory of evaporation may be necessary. We will therefore be most interested in cases where the large majority of particle production occurs while $T_H < M_*$.
 
Assuming $g_L = g_{\overline{L}}$, $\delta N_\bl$ is given explicitly by

\begin{equation}
     \delta N_{\bl} = \int_{M_f}^{M_\pbh} \frac{1}{T_H} G(\mu/T_H) dM,
\end{equation}
where we have defined

\begin{equation}
    G(\mu/T_H) \equiv \frac{g_L \Big(\li_3(-e^{\mu/T_H}) - \li_3(-e^{-\mu/T_H}) \Big)}{\frac{7 g_{other} \pi^4}{240} - 3 g_L \Big(\li_4(-e^{-\mu/T_H}) + \li_4(-e^{\mu/T}) \Big) }
\end{equation}
As we will see shortly, we are interested in $\delta N_{\bl}$ for particular values of $M_*$. We can understand the dependence of $\delta N_{\bl}$ on $M_*$ in the following way: the function $G$ is peaked when $|\frac{\mu}{T_H}| \approx 4.5$. Since $T_H = M_p^2/M$, $\delta N_{\bl} \sim \int M dM \sim M_{\mathrm{peak}}^2$, where $M_{\mathrm{peak}}$ is the mass of the black hole at the time of peak production. Since $\mu/T_H \sim (\frac{M_p}{M_*})^4 (\frac{M_p}{M})^6$ for an uncharged black hole, $M_{\mathrm{peak}} \propto M_*^{-2/3}$ and $\delta N_{\bl} \propto M_*^{-4/3}$. 

Lastly, we note the assumptions used for the initial PBH mass spectrum. Because the large majority of baryons are produced near $M \sim M_{\mathrm{peak}}$, as long as the initial mass of the PBHs is greater than this value ($M_\pbh > M_{\mathrm{peak}}$) and PBHs have sufficient time to evaporate, the baryon abundance is nearly independent of the initial PBH mass. For the parameter space explored in this paper, we find that $M_{p} < M_{\mathrm{peak}} < 10^5 M_p$. Since CMB constraints dictate that $M_\pbh \geq 10^5 M_{p}$ \cite{fujita2014, planckcollaboration2020}, we are justified in assuming that the PBHs considered here will always start out with $M > M_{\mathrm{peak}}$. Additionally, in order for sphaleron processes to be efficient, we require $M_\pbh \lesssim 10^{11} \hspace{1mm} M_p$ \cite{hamadaBaryonAsymmetryPrimordial2017}.Thus, we assume $10^5 M_{p} < M_\pbh < 10^{11} \hspace{1mm} M_p$ throughout, stressing again that the baryon yield is independent of the initial mass for any value in this range. 


\subsection{Results}
\label{sub:relicresults}

If a PBH ceases to evaporate at or below the Planck scale, the population of PBHs responsible for the baryon asymmetry will constitute part or all of the cosmological dark matter abundance as the relic PBHs behave precisely as cold dark matter \cite{Lehmann:2021ijf}. If this is the case, obtaining the correct asymmetry ($Y_{\bl} \equiv \frac{n_{\bl} - \overline{n}_{\bl}}{s}  \sim 8.6 \times 10^{-11}$ \cite{particledatagroup2018}) while simultaneously satisfying $\Omega_{\mathrm{CDM0}} = \Omega_{\mathrm{PBH0}}$ gives a unique prediction for the number density, or equivalently, the final mass of the PBHs for a given value of $M_*$. To see this, consider

\begin{equation}
\label{eq:dnl}
      8.6 \times 10^{-11} \sim \frac{n_{\bl}}{s} = \frac{n_{\pbh}}{s} \delta N_\bl(M_*)
\end{equation}
where $M_f$ is the final mass of each PBH when it has ceased evaporating and $Y_{\mathrm{BH}} \equiv \frac{n_{\pbh}}{s} \sim 1.8 \times 10^{-28} \Big( \frac{M_p}{M_f}\Big)$ is a constant fixed by $\Omega_{\mathrm{CDM0}}$, assuming that the entirety of the dark matter is comprised of relic black holes. We note that there is a maximum value of $Y_{\mathrm{BH}}$ determined by the energy density of the universe comprised of PBHs at the time of formation given by $Y_{\mathrm{BH, max}} \sim \frac{T_\gamma}{M_{\mathrm{PBH}}}$, where $T_\gamma \sim 10^{11} \mathrm{GeV} (\frac{10^5 M_p}{M_{\mathrm{PBH}}})^{3/2}$ \cite{carr2020}. For $M_* \geq 10^{-6} M_p$, peak production occurs at $M_\mathrm{peak} \lesssim 10^5 M_p$, the lowest initial PBH mass allowed by CMB constraints. For self-consistency, we therefore evaluate $Y_{\mathrm{BH, max}}$ at $M_\mathrm{PBH} = 10^5 M_p$ for $M_* \geq 10^{-6} M_p$. Throughout the paper, we will require $Y_{\mathrm{BH}} < Y_{\mathrm{BH, max}}$. 

\begin{figure}
    \centering
    \includegraphics[scale=0.6]{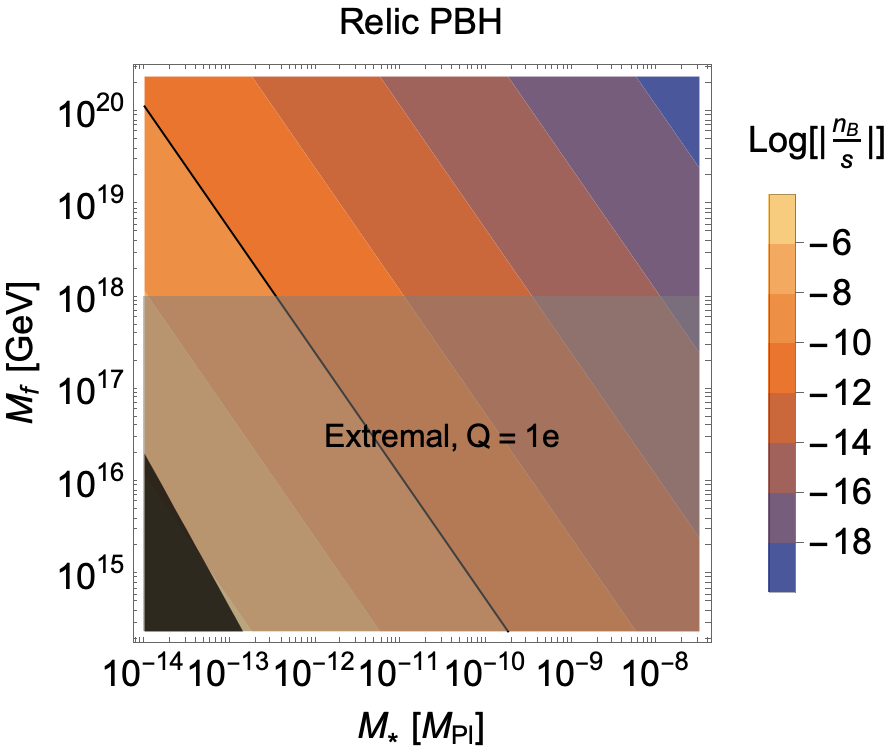}
    \caption{Final mass of Planckian relics. The correct baryon asymmetry corresponds to the black line. Peak production occurs while $T_H > M_*$ in all regions shown in this figure (see text for details). The shaded region indicates where a relic black hole with charge $Q = 1e$ is extremal. The black shaded region indicates where $Y_{\mathrm{BH}}$ is too small to fit the observed DM relic abundance. The lowest value of $M_*$ shown ($\sim 10^{-14} M_p$) corresponds to $M_{\mathrm{peak}} \sim 10^{11} M_p$, the sphaleron bound. For values below this, the black holes would not asymmetrically evaporate efficiently until after sphaleron processes freeze out (see the last paragraph of \ref{sub:BLCharge} for details).}
    \label{fig:Mf}
\end{figure}

For some value of $M_*$, we find the value of $\delta N_\bl$ needed to obtain the correct $Y_{\bl}$. Plugging this into Eq.~(\ref{eq:dnl}) we can uniquely determine $M_f$ as shown in Fig. \ref{fig:Mf}. We see that the correct baryon abundance can be obtained as indicated by the solid black contour, provided $M_*$ is sufficiently small. As explained above, $\delta N_{\bl} \propto M_*^{-4/3}$, and thus $M_*$ must be $\lesssim 10^{-12} M_p$ in order to have a non-extremal charged relic black hole and obtain the observed baryon asymmetry. In this case, $T_H > M_*$ during peak baryon production, meaning the effective field theory may break down. The region where a relic black hole with charge $Q = 1e$ becomes extremal is indicated by the gray shading and is given by $M_f \leq \sqrt{8 \pi} Q M_p \approx \times 10^{18} \hspace{1mm} \mathrm{GeV}$. The prospect of direct detection disappears in this region, as black holes with this mass cannot bear electric charge\footnote{The black holes could be charged under some other unbroken $U(1)_{\mathrm{dark}}$ symmetry, altering the metric in the same way as electric charge. This could create a different scale at which black holes become extremal with respect to the $U(1)_{\mathrm{dark}}$ charge. Here, we limit our discussion to electrically charged black holes and refer the reader to \cite{bai2020} for more detail.}. Of course, uncharged relics could exist in this otherwise extremal region. However, this is somewhat of a nightmare scenario since such a DM candidate would evade all current avenues of detection.

\section{Asymmetric Dark Matter}
\label{sec:Asymmetric}


The fact that $\Omega_{\mathrm{CDM}}$ is on the same order as $\Omega_B$ suggests that a similar formation mechanism may be responsible both for the abundance of baryons and dark matter. Here, we examine the possibility that the relic DM abundance is due entirely to the asymmetric component of a DM species and assume that PBH  evaporation is complete, leaving no relics.

\subsection{Theory}

An evaporating black hole will preferentially radiate DM over anti-DM through the same mechanism as \ref{eq:evap}. Allowing for a different scale of the theory, we have a new coupling

\begin{equation}
   \mathcal{S} \supset \frac{1}{M_{*\chi}^4} \partial_\alpha (\mathcal{R}_{\mu \nu \rho \sigma} \mathcal{R}^{\mu \nu \rho \sigma})J_{\mathrm{\chi}}^\alpha,
\end{equation}
where $M_{*\chi}$ is the equivalent of $M_*$ in the dark sector; it defines the energy cutoff scale of the effective field theory. 

As before, a chemical potential is generated via the evaporation of the black hole. We take the dark degrees of freedom to be $g_\chi = \overline{g_\chi} = 2$, noting that allowing for a large number of degrees of freedom in the dark sector does not significantly change our results (see e.g. \cite{morrison2019}).

For a given value of $M_*$, the requirement of producing the correct abundance of baryons fixes the abundance of black holes during the period of particle production

\begin{equation}
    Y_{\mathrm{BH}}(M_*) \simeq \frac{8.6 \times 10^{-11}}{\delta N_{\bl}(M_*)}.
\end{equation}
Since the same black holes are responsible for the production of the asymmetric dark matter, we can make a unique prediction for the dark matter mass as follows. The net yield of dark matter particles produced is $n_\chi/s = Y_{\mathrm{BH}} \delta N_\chi(M_{*\chi})$, where $\delta N_\chi$ is the number of asymmetric DM particles produced per black hole. The relative abundance of dark matter is 

\begin{equation}
\label{eq:omegachi}
    \Omega_\chi = \frac{n_\chi M_\chi}{\rho_c}.
\end{equation}
Therefore, the mass of the dark matter particle giving the observed cosmological DM abundance is

\begin{equation}
    M_\chi = \frac{\rho_c \Omega_{\mathrm{CDM}}}{s Y_{\mathrm{BH}}(M_*) \delta N_\chi(M_{*\chi})}.
\end{equation}
We note that the net yield of dark matter particles discussed here is what is left over after the annihilation of the bulk of the DM with anti-DM. The total amount of dark matter produced initially is generically very large, allowing for even a small asymmetry to lead to the current cosmological abundance of dark matter after annihilation. To illustrate this, we follow \cite{morrison2019} which finds that the total amount of DM initially produced (both $\chi$ and $\overline{\chi}$) is given as

\begin{equation}
    \begin{split}
    \Omega_{\mathrm{DMi}} \approx 6.5 \times 10^7 \beta \hspace{1 mm} \Big(\frac{M_\chi}{\mathrm{GeV}}\Big) \Big( \frac{M_\pbh}{M_{p}}\Big)^{1/2}, \hspace{3 mm} M_\chi < M_{p}^2/M_\pbh     \\
    \Omega_{\mathrm{DMi}} \approx 6.5 \times 10^7 \beta \hspace{1 mm} \Big(\frac{M_\chi}{\mathrm{GeV}}\Big) \Big( \frac{M_{p}^7}{M_\pbh^3 M_\chi^4}\Big)^{1/2}, \hspace{3 mm} M_\chi > M_{p}^2/M_\pbh,
    \end{split}
\end{equation}
where $\beta \equiv M_\pbh \frac{n_\pbh(T_i)}{\rho_\mathrm{rad}(T_i)}$ is the relative density of PBH to radiation at the time of PBH formation with temperature $T_i$ (assuming the PBHs do not lead to an early period of matter domination). The temperature at formation is given approximately by 

\begin{equation}
    \frac{T_i}{M_{p}} \approx 0.87 \Big(\frac{M_{p}}{M_\pbh}\Big)^{1/2}.
\end{equation}
We verify that $\Omega_{\mathrm{DMi}} >> \Omega_{\mathrm{CDM}}$ for the range of $M_\pbh$ and $M_\chi$ considered in this work. Thus, in analogy to the baryonic asymmetry, we consider the scenario in which the vast majority of DM coannihilates. The asymmetric component remains and constitutes the cold dark matter today. 

\subsection{Constraints}
\label{sec:asymconstraints}
We consider constraints from partial wave unitarity, which comes from requiring that the probability of inelastic scattering is not greater than $1$ \cite{griest1990}. We note that asymmetric DM requires a larger annihilation cross section than does symmetric DM, leading to a tighter upper bound on $M_\chi$ than in a purely symmetric DM scenario \cite{baldes_asymmetric_2017}. Assuming that the S-wave channel is the dominant contribution and that the vast majority of $\chi$ and $\overline{\chi}$ coannihilate, this bound corresponds to ($M_\chi \lesssim 30 \times 10^3 \hspace{1 mm} \mathrm{GeV}$).

We also consider constraints due to the capture of asymmetric DM by neutron stars. It has been shown that an asymmetric component of dark matter could accumulate at the center of neutron stars, eventually collapsing into a black hole and rapidly destroying its host neutron star. This places limits on the scattering cross section with nucleons for asymmetric dark matter of a given mass \cite{Kouvaris2010,Kouvaris2011,mcdermott2012,Garani2019,Dasgupta2020}. Observations of nearby pulsars can place constraints on the asymmetric dark matter mass $M_{\chi}$, assuming a particular scattering cross section. However, nearby pulsars provide weaker constraints on the mass of fermionic DM than constraints from partial wave unitarity \cite{Garani2019}; assuming a cross section of $10^{-45} \ \mathrm{cm^2}$, the most stringent bound due to this constraint is $M_{\chi} \lesssim 10^7 \ \mathrm{GeV}$.


It is possible for this constraint to be affected by DM self interactions and the capture rate of asymmetric DM in neutron stars. Even with DM self interactions, DM masses smaller than $10^6 \ \mathrm{GeV}$ cannot induce the collapse of neutron stars \cite{2019Gresham}. Detailed calculations of the capture rate \cite{2021Anzuini,2021Bell2} show that the capture rate can be reduced by up to 3 orders of magnitude when nucleon structure and strong interactions are accounted for. Since $M_{\chi}$ is inversely proportional to the capture rate for $M_{\chi} \gtrsim 1 \ \mathrm{GeV}$ \cite{Garani2019}, the reduction in the capture rate relaxes this constraint. Therefore, we take the unitarity limit to be the most conservative upper-bound on the asymmetric DM mass. 


There also exist constraints on how warm the DM produced via evaporation can be. The current velocity of particle dark matter produced from the evaporation of a black hole is \cite{lennon_black_2018, fujita2014}

\begin{equation}
    v_\chi = \frac{p_{\mathrm{now}}}{M_\chi} \approx 4 \times 10^{-31} \Big(\frac{M_\chi}{M_{p}}\Big)^{-1} \Big(\frac{M_\pbh}{M_{p}}\Big)^{1/2}.
\end{equation}
Comparing this to present constraints on the velocity of relic dark matter ($v_\chi \lesssim 4 \times 10^{-8}$) \cite{baldes2020}, one obtains an upper limit on the dark matter mass \cite{lennon_black_2018, fujita2014}

\begin{equation}
    \frac{M_\chi}{1 \mathrm{GeV}} \gtrsim 3 \times 10^{-5} \Big( \frac{M_\pbh}{M_{p}}\Big)^{1/2}.
\end{equation}
The smallest possible initial PBH mass allowed by CMB constraints \cite{planckcollaboration2020} ($M_{\mathrm{PBH}} \sim 10^5 M_{p}$) gives the smallest possible $M_\chi$ for DM produced through Hawking evaporation

\begin{equation}
    M_{\chi} \gtrsim 10 \hspace{1 mm} \mathrm{MeV},
\end{equation}
which is in good agreement with limits derived from cutoffs in the linear matter power spectrum \cite{baldes2020}. 

We note that these limits depend on the model of asymmetric DM. In the simplest instances, the portion of the dark matter from asymmetric dark matter will likely fall into thermal equilibrium due to the elastic scattering off of the thermal bath induced by the same interactions that deplete the symmetric component; however, this is not necessarily the case in {\em any} model, since it involves a non-trivial crossing symmetry between the pair-annihilation processes and the cross-symmetric elastic scattering process. With this consideration, the overall viable asymmetric dark matter mass range considered here is $10 \hspace{1 mm} \mathrm{MeV} \leq M_\chi \leq 30 \times 10^3 \hspace{1 mm} \mathrm{GeV}$.

\subsection{Results}

\begin{figure}
    \centering
    \includegraphics[scale=0.6]{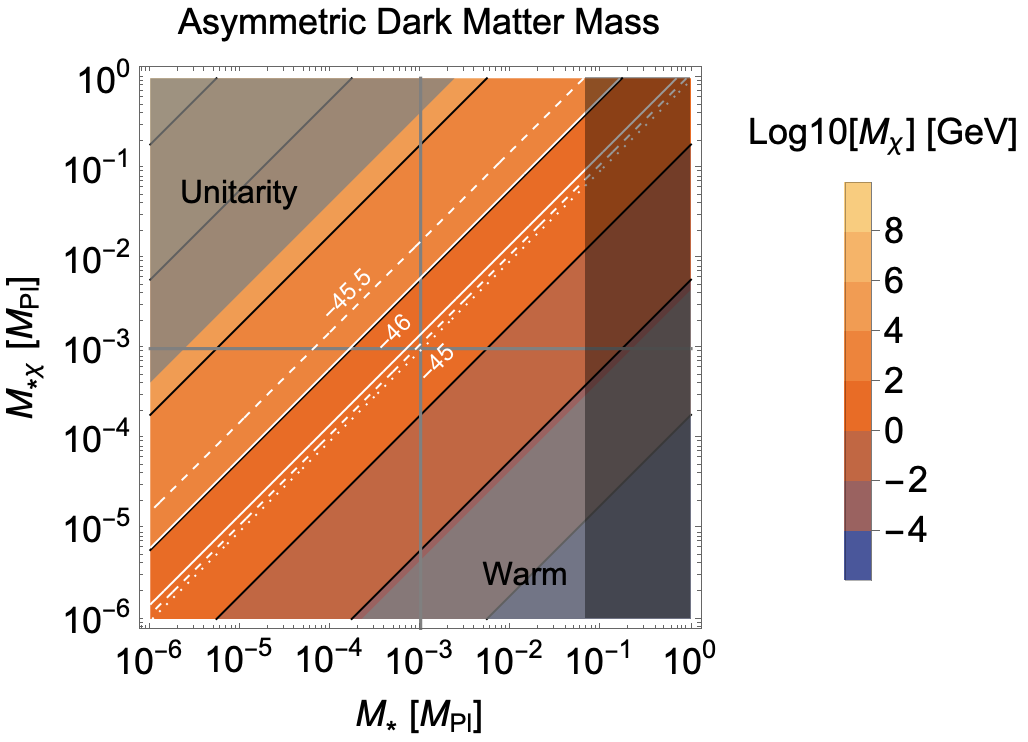}
    \caption{Mass of asymmetric DM particle assuming $\Omega_{\mathrm{CDM}} = \Omega_{\chi}$. An upper bound on the DM mass from partial wave unitarity is indicated by the gray shaded region in the upper left corner of the figure. A lower bound from the dark matter being too warm is shown in the lower right corner. For $M_* \gtrsim 0.07 M_p$, the correct baryon asymmetry cannot be obtained without $Y_{\mathrm{BH}} > Y_{\mathrm{BH, max}}$, indicated by the dark shaded region on the right of the figure, (see Sec.  \ref{sub:relicresults}). Upper limits on the DM-nucleon cross section (in units of $\mathrm{cm}^2$) from the XENON1T experiment for a given $M_{\chi}$ are shown in white with $\mathrm{log_{10}}\sigma = -46, -45.5, -45$ corresponding to the solid, dashed, and dotted lines, respectively \cite{Aprile2018}.}
    \label{fig:Asymmetric}
\end{figure}

Contours of constant particle dark matter mass producing the observed cosmological dark matter abundance is shown in Fig.~\ref{fig:Asymmetric} as a function of $M_*$ and $M_{*\chi}$. The dashed line indicates where $M_* = M_{*\chi}$, which is by no means required, but is perhaps a natural choice of parameters. The blue lines indicate the regions below which peak production occurs while $T_H > M_*, M_{*\chi}$; the upper right quadrant is where the effective field theory description is most appropriate. The white contours represent the upper limits on the DM-nucleon cross sections for particular values of $M_{\chi}$, given the direct detection constraints from the XENON1T experiment \cite{Aprile2018}. The contours for $\mathrm{log_{10}} \sigma = -46$ (with $\sigma$ in units of cm$^2$ and log in base 10 implied hereafter) and $\mathrm{log} \sigma = -45.5$ show up in pairs, with the innermost (solid) contours corresponding to the lower cross section. There is only a single contour (dotted) for $\mathrm{log} \sigma = -45$, which appears on the bottom right. 

Some of the relevant parameter space in this scenario is constrained due to partial wave unitarity, as indicated by the gray shading in the upper left region of the plot. We find that the constraints from unitarity are stronger than those from pulsar capture for the entire region of interest and therefore omit the capture bounds in this figure. The lower warm DM bound is indicated by the gray shading in the lower right corner of the figure.

There is a significant window in the upper right quadrant of the figure where this scenario is not ruled out by unitarity, pulsar capture, or warm DM constraints and the EFT is a valid description during peak production of both baryons and DM. This demonstrates that the baryon asymmetry and the cosmological DM abundance can simultaneously be satisfied by nearly identical production mechanisms. 

For completeness, one may want to consider the possibility that a significant population of relic black holes is left after the production of the asymmetric DM. Thus we are motivated to introduce a new two-component DM scenario in the following section.

\section{Two Component DM}
\label{sec:TwoComponent}

\subsection{Introduction}


We will now see that if the dark matter is comprised of both a particle \textit{and} a population of relic black holes, baryogenesis and the dark matter abundance can both be satisfied, all while maintaining a valid effective field theory and allowing for the cessation of BH evaporation near or below the Planck scale.

\subsection{Theory}

In this two-component dark matter scenario, the particle $\chi$ is produced in the same manner as in section \ref{sec:Asymmetric}; however, now, we additionally assume that black hole evaporation ceases at some final black hole relic mass $M_f$. This introduces a new free parameter which we parameterize as

\begin{equation}
    M_f = 10^\eta M_\chi.
\end{equation}

\begin{figure}
    \centering
    \includegraphics[scale=0.7]{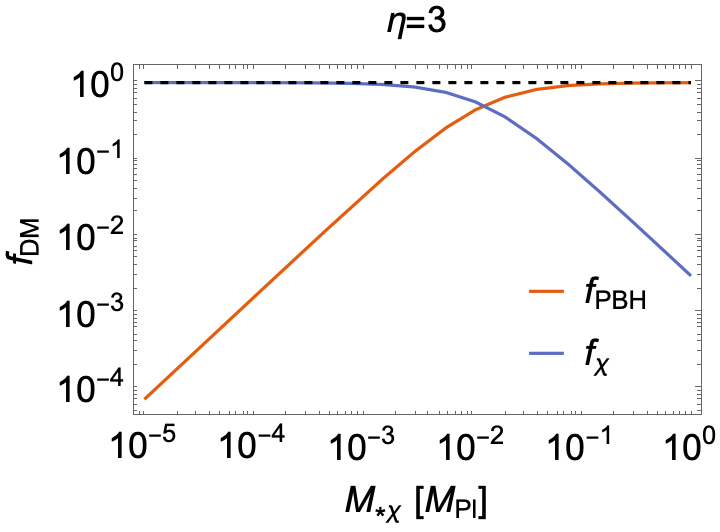}
    \caption{Fraction of DM comprised of PBHs and $\chi$ particles for $\eta = 3$. The fraction of each is determined by how many $\chi$ particles are produced per black hole, $\delta N_{\chi}(M_{*\chi})$.}
    \label{fig:fDM}
\end{figure}

Since the black holes do not evaporate away, their number is now conserved. The relative abundance of black holes is then
\begin{equation}
\label{eq:omegaPBH}
    \Omega_{\mathrm{PBH}} = M_f Y_{\mathrm{BH}}\frac{s}{\rho_c}. 
\end{equation}
In this case, the mass of the $\chi$ particle required in order to obtain the correct dark matter abundance is

\begin{equation}
    M_\chi = \frac{\rho_c \Omega_{\mathrm{CDM}}}{s Y_{\mathrm{BH}}(M_*) (\delta N_\chi(M_{*\chi}) + 10^\eta)}.
\end{equation}
The fraction of DM comprised of each component is determined by the number of $\chi$ particles produced per PBH, which is solely a function of $M_{*\chi}$. This relationship is shown explicitly in Fig. \ref{fig:fDM}.

\subsection{Constraints}

We consider the same constraints as in \ref{sec:asymconstraints}. However, in the two component scenario we address the fact that the DM particles do not generally constitute the entire dark matter abundance for an arbitrary value of $M_{*\chi}$. In this case, we scale the number of DM particles by the fraction of dark matter made up of the asymmetric DM component
\begin{equation}
   f_\chi \equiv \frac{\Omega_{\chi}}{\Omega_{\mathrm{CDM}}} = \frac{\Omega_{\chi}}{\Omega_{\chi}+\Omega_{\mathrm{PBH}}} = \left(1+\frac{10^{\eta}}{\delta N_{\chi}(M_{*\chi})} \right)^{-1},
\end{equation}
considering that $\Omega_{\chi}$ and $\Omega_{\mathrm{PBH}}$ are given by Eqs. \ref{eq:omegachi} and \ref{eq:omegaPBH} respectively. Thus, the constraints on asymmetric DM from partial wave unitarity and capture by neutron stars are scaled corresponding to the fraction of DM made up of this component. With this scaling, the constraint due to capture by neutron stars becomes even weaker. 

The limit on the asymmetric DM mass from unitarity becomes stronger as $f_\chi$ becomes smaller. This is because a larger cross section is required to suppress the abundance, which pushes the limit to smaller $M_\chi$. Since $\Omega_\chi \sim (\sigma v_\mathrm{rel})$, the maximum cross section allowed by unitarity for S-wave annihilation ($\sigma v_\mathrm{rel} \sim M_\chi^{-2}$) implies the maximum allowed $M_\chi$ scales as \cite{baldes_asymmetric_2017}

\begin{equation}
    M_\chi \lesssim 30 \times 10^3 \hspace{1 mm} f_\chi^{1/2} \hspace{1 mm} \mathrm{GeV}.
\end{equation}
Note, however, that this limit only applies to the asymmetric component of the dark matter. The relic black hole population can still account for the total dark matter abundance in regions where asymmetric DM is constrained by unitarity, provided $f_\mathrm{PBH} \sim 1$.

\subsection{Results}
\label{sec:results}
\begin{figure}
    \centering
    \includegraphics[scale=0.55]{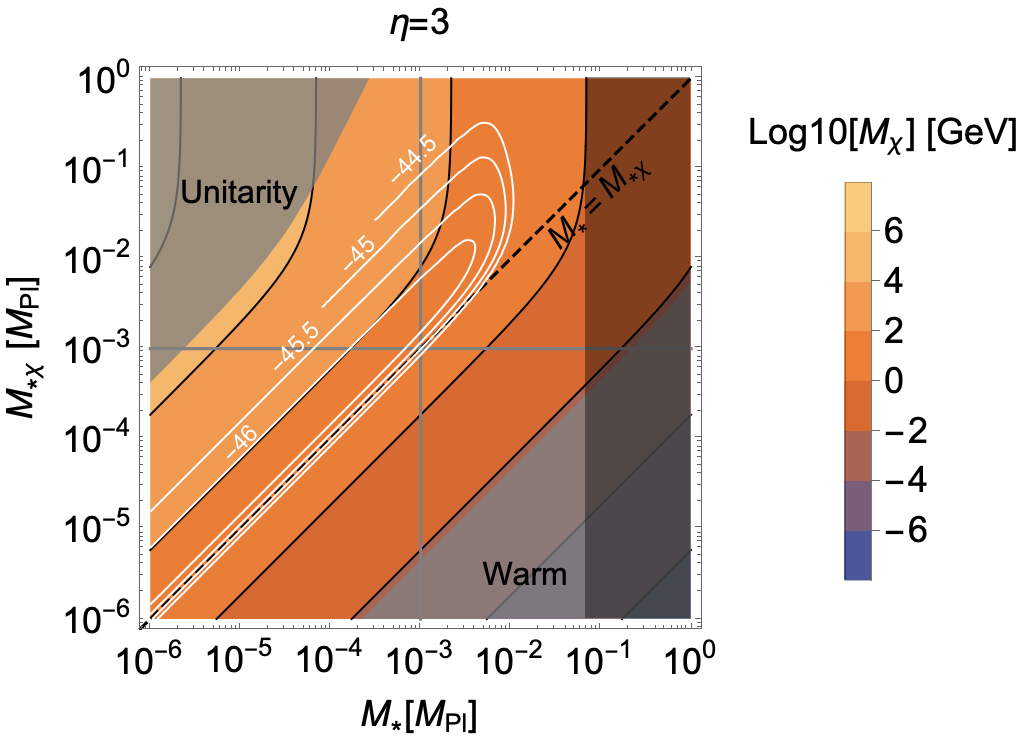}
    \caption{Mass of Asymmetric DM particle in the two-component DM scenario for $\eta = 3$.}
    \label{fig:TwoComponentEta3}
\end{figure}

\begin{figure}
    \centering
    \includegraphics[scale=0.5]{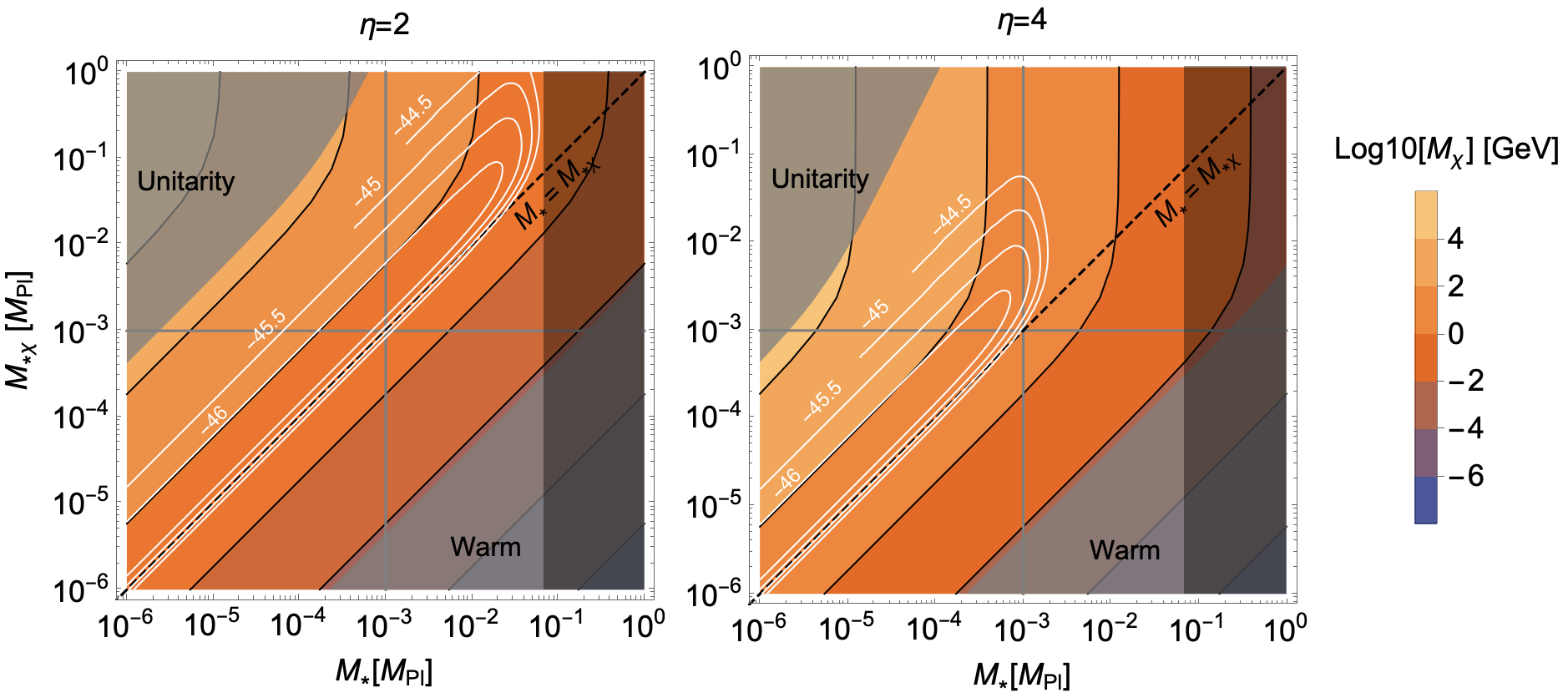}
    \caption{Mass of Asymmetric DM particle in the two-component DM scenario for $\eta = 2,4$.}
    \label{fig:TwoComponentEta24}
\end{figure}

We show the resulting mass of the dark matter particle as a function of $M_*$ and $M_{*\chi}$ for the case of $\eta = 3$ in Fig. \ref{fig:TwoComponentEta3} and for $\eta = 2,4$ in Fig. \ref{fig:TwoComponentEta24}. As before, the dashed line corresponds to $M_* = M_{*\chi}$ and the upper right quadrant indicates the region where $T_H < M_*, M_{*\chi}$ for peak production. The upper left shaded region indicates constraints from partial wave unitarity. Since the constraints from pulsar destruction are less stringent than those from unitarity, they do not appear in the plots of Figs. \ref{fig:TwoComponentEta3} and \ref{fig:TwoComponentEta24}. In the lower right shaded region, the dark matter produced through evaporation would be too warm today. As before, the white contours represent the upper limits on the DM-nucleon cross sections for particular values of $M_{\chi}$ from XENON1T. Since this constraint only applies to the asymmetric DM component, we scale the limits by $\sigma \to \sigma/f_{\chi}$. 

The shape of the contours can be understood as follows. In the upper region of the plot where the contours are nearly vertical, the abundance of dark matter particles that is produced is very low, meaning that the relic black holes account for the vast majority of the dark matter. That is why the contours are independent of $M_{*\chi}$ in this region. But at a certain point, the production of dark matter particles becomes large enough that now they dominate the contribution to the totality of dark matter. At the point where the contours turn over, the total contribution is equal from both components. Intriguingly, this equal-component scenario can occur for $M_* = M_{*\chi}$ in the region where the EFT is still valid and is completely unconstrained by partial wave unitarity, neutron star capture, and warm dark matter limits. The possibility of detecting a signal from this region of parameter space hinges on possible model-dependent direct or indirect detection signals from the asymmetric dark matter component and, possibly, for sufficiently heavy PBH evaporation relics (thus for large-enough $\eta$, in fact larger than what shown in the plots), direct detection of such relics if charged \cite{lehmannDirectDetectionPrimordial2019}.

\section{Discussion and Conclusions}
\label{sec:discussion}

We have shown that a population of PBHs could have produced the observed baryon asymmetry through gravitational baryogenesis, while simultaneously accounting for the dark matter abundance in the universe assuming evaporation at or below the Planck scale. We also investigated an alternative scenario in which dark matter is entirely composed of a DM particle species, generated asymmetrically in a manner analogous to the mechanism that produces the baryon asymmetry. We showed that an extremely large mass is necessary for the DM particle to constitute $100 \%$ of the dark matter abundance while satisfying baryogenesis. In both cases (with or without an asymmetric dark matter component), the effective operator scale is well below the Hawking temperature of the evaporating black hole, signalling a possible breakdown of the validity of the effective theory description at energies relevant for the production of the desired asymmetry. 

We therefore also considered the possibility that dark matter is comprised of a relic PBH component and of an asymmetric DM component stemming from gravitational genesis, similar to the baryonic matter. We demonstrated that this model satisfies both the dark matter abundance and baryogenesis for a reasonable dark matter mass, and it is possible for there to be an equal contribution from the two DM components in an unconstrained region of the parameter space where EFT is applicable. This scenario can, importantly, also easily accommodate the requirement for the validity of the effective theory, $T_H>M_*,\ M_{*\chi}$.

Depending on the value of $\eta$, the mass of the evaporation relics relative to the asymmetric dark matter mass, the relics might well be too light to carry any charge because of extremality, in which case they would not be detectable. The detection of the asymmetric dark matter component is possible, but is entirely model-dependent (here, we calculated the required scattering rate off of protons for the dark matter to be detectable). Depending on the relic symmetric component as well as on possible additional interactions, indirect detection is also possible. Since all of these possibilities are generic of any asymmetric dark matter model, and not specific to the production mechanism we discuss here, we refer the Reader to recent reviews on the topic, such as Refs.~\cite{Petraki:2013wwa, Zurek:2013wia}.

\bibliographystyle{JHEP}
\bibliography{main}

\end{document}